# Local Structure of Functional Solids

Shin-ichi Shamoto[1]

[1]*Advanced Science Research Center, Japan Atomic Energy Agency,*

*Tokai, Ibaraki 319-1195, Japan*

The local structure study reveals important aspects of the physical properties, because it is closely related to the electronic structure. Standard crystallographic analysis based on a space group fails to observe disorder in the crystal structure. Many functional and industrial materials have disorder, because of the optimized modifications of crystal structures such as the substitution of an element. Doped elements inevitably induce local strains in the matrix. Some neutron total scattering results are reviewed to show the local functional properties of solids from optical recording to negative thermal expansion to Mott transition materials.

# 1. Introduction
*1.1 Disorder*

    A substance usually has three states of matter, namely gas, liquid, and solid. It may also be a liquid crystal. In reality, it can be between these states under some conditions. These states may coexist as a phase separation in the first-order phase transition. Each phase may have a different local structure, which results in multiple different crystal structures even in a single compound. It may have both amorphous and crystalline phases in a material. The same phase separation with nonuniform compositions may be observed even in an amorphous phase as a result of highly nonequilibrium material processing. The simplest case of a noncrystalline part is a defect in a material. The defect is inevitably created beyond the activation energy during a thermal equilibrium at a finite temperature. The number of defects increases not only with increasing temperature but also with increasing consistent element number. These ordered and disordered states may have individual characteristic physical properties. They may be macroscopically separated similarly to water and ice because of the density difference between them. On the other hand, composite materials including artificial lattice films may be composed of various materials in different forms and length scales. Even a single phase can be composed of various structures such as layered, tunnel, and cage local structures.[1] These materials have interfaces between these local structured parts and states. Interestingly, this discussion is not limited to their crystal structures, but also to their electronic states. Their electronic states may include metallic, semimetallic, semiconductive, insulating, and/or magnetic states. They may be charge- and/or orbital-ordered states. One of them may be a ferroelectric and/or ferromagnetic state. At the same time, they may include the other states such as antiferromagnetic, spin liquid,

spin glass, and non-collinear magnetic states. These multiple properties can be shared in a single material, whereas composite materials may also share these properties in their individual parts. A perfect crystal is simple and beautiful, whereas a disordered material can be attractive and intriguing to study because of its complexity and rich properties, similarly to a living creature. These materials exhibit various responses in relation to their physical properties, corresponding to their local structures. Here, some typical functional disorder will be reviewed.

*1.2 Chemical bond and electronic structure*

Bond length is a characteristic property of a chemical bond. It depends on the number of electrons of specific chemical bonds described as the bond order that is half the number of electrons in the bonding orbitals subtracted by half the number of electrons in the antibonding orbitals. In the case of a covalent carbon-carbon bond, one may find a large change in the bond length with the bond order. Typically, the length of a single bond is 1.54 Å with bond order $p=1$ (1$\sigma$ bond). It becomes 1.34 Å for a double bond with $p=2$ (1$\sigma$ bond and 1$\pi$ bond). It is further decreased to 1.20 Å for a triple bond with $p=3$ (1$\sigma$ bond and 2$\pi$ bonds). The relative change exceeds 10% of the bond length. In covalent $\sigma$ bonding, the difference in the bond length per electron becomes large, suggesting large electron-lattice coupling. The change in the bond length also depends on the orbital. One of the strongest chemical bonds is a carbon 2$p$ orbital bond, which has a change in the bond length of about 0.10 Å/e from a single bond to a double bond. With increasing energy difference between two bonding orbitals, the heteropolarity increases. According to the laws of quantum mechanics, the charge transfer cannot reach unity even in a highly ionic bond because of the finite exchange integral between two neighboring orbitals. The non-integer charge transfer is consistent with the experimental results of valence numbers based on the bond valence sum.[2]

A characteristic physical property affected by the bond length in materials is the electronic band width. A small amount of a doped element could change the physical properties dramatically if the material was close to a critical point of the phase transition. The Mott transition is one example, where the bond length is also expected to show a large change because of the dramatic transition of the electronic states.

The most dramatic change in a chemical bond is a chemical reaction in a liquid. Let's think about a simple heteropolar diatomic molecule consisting of cation A and anion B in a solution, where the highest occupied molecular orbital (HOMO) is the bonding orbital and the lowest unoccupied molecular orbital (LUMO) is the antibonding orbital.[3] Then, there may be a charge transfer from the cation (A) to the anion (B) in the diatomic molecule. If a donor in the solution approaches anion B, the bond length between A and B will be increased because of the charge transfer from the donor to the LUMO with the antibonding character. In the opposite case, if an acceptor approaches cation A, the bond length will also be increased because of the charge transfer to the acceptor from the HOMO with the bonding character. At a large charge transfer, the chemical bond of the heteropolar diatomic molecule will be cleaved. This type of solute and solvent interaction changes the electron density distribution in a molecule, resulting in bond length variation or chemical reactivity.[4] The bond length variation may not be limited only in the molecule, but may also occur in a solid.

For a solid, the observed chemical bond helps us to understand the electronic state because of the close intrinsic correlation between them. One of the recent advancements in physics is the understanding of the Mott transition. Photoemission spectroscopy is a powerful tool for observing the Mott transition. According to Ref. 5, the electronic structure of a Mott insulator can be well expressed by a cluster model with a transition metal ion and ligand anions. It is based on ligand field theory for a cluster. This idea is consistent with the localized electron picture for a cluster. According to the ligand field theory, the electrons in the transition metal $d$ orbital always belong to the antibonding orbital state, whereas the bonding orbital electrons are at the anion $p$ orbitals. Because the electrons are localized in each cluster, the energy dispersion becomes flat in the momentum space. If the system is in a magnetically ordered state, the electronic state also depends on the magnetic correlation. The charge transfer integral depends on the neighboring magnetic spin correlation. If it is in an antiferromagnetic (AF) state, two polarized up and down spin bands become staggered at the neighboring sites. This situation reduces the effective electron hopping between neighboring sites, suggesting an increased bond length. The details will be discussed in sect. 3.4 on the Mott transition.

**2.   Measurement and Analysis Techniques**

Local structure can be described by local bond length and local bond angle, which may vary in place and time. Scattering is basically the interference between two waves scattered from two atoms. It may be enhanced or not, depending on the phase difference. A powder diffraction pattern results from the summation of these scattered intensities after averaging over the solid angles for all pairs of atoms. Because of the characteristics, it is easy to measure bond lengths between two individual atoms, whereas bond angle estimation is difficult, which requires the other bond lengths and the surrounding atomic configuration. The estimated bond angle may sometimes be unique and sometimes model-dependent.

There are some methods of measuring the local structure in a crystalline material. One of them is the total scattering of neutrons and X-rays, except for real space imaging methods such as transmission electron and scanning tunneling microscopes. Another is the extended X-ray absorption fine structure (EXAFS), which is also known as a local structural study tool.[6] This method is very sensitive to the relative change in bond length. However, the $r$-range is usually limited in a short range. Holography is also known as the local structure probe.[7]

The total scattering is a simple tool to study the local structure. The analysis is based on the atomic pair distribution function (PDF), $g(r)$, which can be derived by a Fourier transformation of scattering pattern, $S(Q)$, over a wide $Q$-range in an absolute scale.[8] A widely used correlation function, the reduced pair distribution function, $G(r)$, is defined in addition to $g(r)$, as

$$G(r) = 4\pi r \rho_0 (g(r) - 1) = \frac{2}{\pi} \int_0^{Q_{max}} Q[S(Q) - 1]\sin(Qr)dQ, \tag{1}$$

where $\rho_0$ is the average number density of the scatterers, the integration $Q$-range is experimentally limited to $Q_{max}$ instead of infinite, and the PDF is defined as

$$g(r) = \frac{1}{\rho_0^2 V} \int_{-\infty}^{\infty} dr' \: \langle \rho(r')\rho(r+r') \rangle , \tag{2}$$

where $\rho(r)$ is the density function as a function of $r$. After the integration in a real space $r$, all the atoms are at $r=0$. This effect corresponds to the loss of phase information in the scattered waves.[9] Our data analysis has been carried out using PDFgetN software.[10] The intensity is usually normalized using a standard vanadium sample. The incoherent intensity is suitable for the normalization. Because of the normalization, it is necessary to measure the sample, the vanadium standard sample, and the used empty cell under identical conditions. In addition, multiple scattering and absorption should be estimated. The Placzek correction is required in a sample with light elements such as hydrogen.[11] This method provides us with a new method of structural study different from the traditional method of crystallographic study based on space groups. There is intrinsic disorder in many of crystalline solids, which cannot be expressed on the basis of a space group. The local structure picture gives us new aspects about some functionalities of existing materials. To understand the meaning of the analysis, a carbon buckyball of $C_{60}$, i.e., a fullerene, is shown as a good example for comparison with the amorphous azafullerene of $C_{59}N$[12] in Fig. 1. Figure 1(a) shows the scattering functions of $C_{60}$ and $C_{59}N$. These scattering functions show similar wavy backgrounds. Figure 1(b) shows their reduced PDFs. They are similar to each other below 8 Å, suggesting a similar molecular structure. The significant difference between them in Fig. 1(a) is the sharp strong Bragg peaks for the face-centered-cubic (fcc) structure of $C_{60}$. This corresponds to the broad peak at about 10 Å in Fig. 1(b).[13] This contrasting feature between $C_{60}$ and $C_{59}N$ is very helpful for understanding what the PDF shows. This ball-type molecule shows a clear size effect in the PDF. In a nanoparticle, crystal structure coherence is similarly limited in a certain real-space range.[14,15]

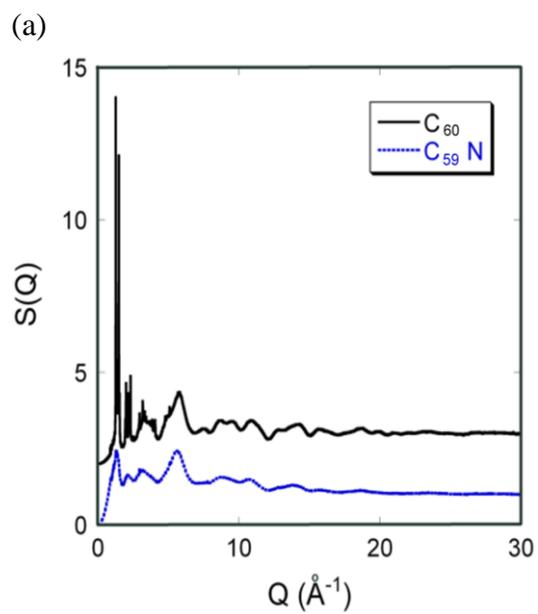

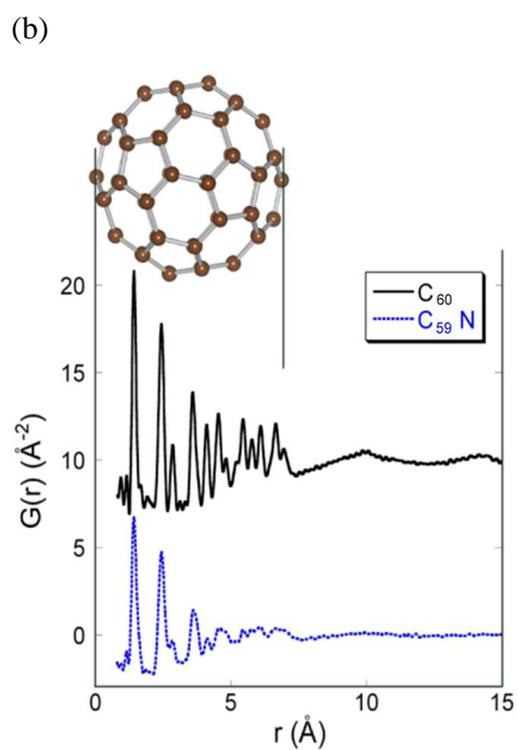

Fig. 1. (Color online) (a) Scattering functions of $C_{60}$ and $C_{59}N$ powder samples. (b) Corresponding reduced atomic pair distribution functions $G(r)$ with a buckyball structure in the same scale.

The coherence length of the crystalline part can be measured using this reduced pair distribution function, $G(r)$, similarly to the fullerene.

Because of the Fourier transformation, the $r$-resolution is determined using the maximum $Q$ value as

$$\Delta r \sim \pi/Q_{max}. \tag{3}$$

On the other hand, the maximum $r$-range is determined by the $Q$-resolution as

$$r_{max} \sim \pi/\Delta Q. \tag{4}$$

The $Q$-resolution mainly depends on the $L_1$ length between a moderator and a sample because of the time-of-flight (TOF) measurement method. The longer $L_1$ length is, the better $Q$-resolution is. At the same time, the intensity becomes weak due to the divergence of neutron beam. In the limit of slow neutrons and backscattering, the $Q$-resolution is inversely proportional to the length $L_1$, whereas the intensity decreases as $1/L_1^2$ with increasing $L_1$. A high $Q$-resolution diffractometer largely loses the neutron intensity. Because of this rule, most diffractometers roughly align on a line in a two-dimensional map of relative intensity and $Q$-resolution depending on the length $L_1$ for one neutron source, although there are other factors such as the moderator, supermirror guide, and focusing devices. From this relationship, it is clear that the simultaneous requirement of both a wide $Q$-range and high $Q$-resolution challenges us to achieve them without compensation of the intensity, time, and sample mass. For example, high-quality statistics can be obtained from a low-$Q$ range, corresponding to a low $r$-resolution in the PDF. Unfortunately, the Fourier transformation analysis requires high statistics of the full diffraction pattern. As a result, this wide $r$-range data only become available by using high-intensity neutron/X-ray diffractometers. Note that the termination errors appear as a wavy ripple in the background. Only the high-intensity data set can lead to small ripples regardless of the $r$-resolution.

For example, the $Q$-resolution of neutron powder diffractometer (NPDF) at LANSCE was about 0.015 Å$^{-1}$ in the $Q$-range from 1.65 to 20.0 Å$^{-1}$, resulting in a real-space range of about 200 Å.[16], which is similar to that of ibaraki materials design diffractometer (iMATERIA) of J-PARC. In the case of high intensity total diffractometer (NOVA) of J-PARC, the real space range becomes about 120 Å, which is similar to that of nanoscale-ordered materials diffractometer (NOMAD) of spallation neutron source (SNS), because of the similar $L_1$ length. By using the wide $r$-space measurement capability, the local structural correlation length can be estimated. Because of the limited $r$-range, the size analysis is still limited to one or two parameters such as the length and distribution width.[17] The local structures of nanomaterials such as surface stress,[18] the surface structure of a gold adsorbent,[19] and water molecule coordination on a surface[20] have been studied.

Depending on the disorder you are interested in, it is important to estimate how high the $r$-resolution should be. In the case of the Jahn—Teller distortion of a transition metal oxide, a typical bond length changes from 1.92 to 2.06 Å. The difference is 0.14 Å, leading to the requirement of a maximum $Q$-value of 25 Å$^{-1}$. In the case of a standard oxide, the maximum $Q$-range is usually limited to 50 Å$^{-1}$, because of the finite lattice vibration even at low temperatures. Therefore, one may expect a 0.07 Å resolution as the highest real-space resolution for this total scattering method. There are exceptions in hard materials with few defects and little distortion. Note that standard crystallographic analysis such as Rietveld analysis has high accuracy for the determination of lattice parameters and atomic positions. The accuracy depends on the $Q$-resolution in addition to the multiplication of lattice parameters by high indices. It may be a good way of estimating a small lattice distortion with high precision by applying this merit even in PDF analysis. For exa mple, there can be a small orthorhombic distortion possibly due to orbital ordering. The orthorhombicity $\varepsilon$ is defined as $\varepsilon=(a-b)/(a+b)$. In a scattering pattern, it is easy to distinguish the structural distortion by measuring the peak splitting related to this distortion. On the other hand, it is difficult to measure the splitting of a PDF peak at a short distance because of the limited $r$-resolution $\Delta r$ in addition to the peak broadening by finite lattice vibration. At a long distance, however, it is possible to find the peak splitting due to the multiplication of a periodic lattice by using a method similar to the reciprocal space analysis.[21] If a domain has a diameter of $n$ unit cells, the orthorhombic distortion can be detected at a long distance under the condition of $\Delta r < n(a-b)$. Under the condition of a large domain with $n$ unit cells, the orthorhombic distortion is averaged in the domain, which becomes similar to the average structural analysis by the Rietveld method, especially with increasing $n$.

## 3.  Research Examples

*3.1 Jahn―Teller distortion*

In a pseudobinary system, for example, the physical properties such as lattice parameters may be regarded as the average of two end materials. The local structure in a PDF pattern clearly shows the individual bond lengths of end materials.[22,23] A typical example is a direct energy gap in a pseudo-binary semiconductor, where the energy gap is determined between the valence and conduction bands at the $\Gamma$ point. Because of the reciprocal point, all orbital energies are averaged. On the other hand, the PDF shows all the individual bond lengths of a material in a pattern. Another good example showing the merit of PDF analysis is the Jahn—Teller distortion in transition metal oxides. The lattice distortion can be observed in bond lengths such as that of Ni-O in $LiNiO_2$.[24] $LiNiO_2$ is a triangular quantum spin system.[25] The crystal structure can be approximated to be the NaCl type. A cation site of the NaCl structure is split into two layers in the (111) plane because of the periodic ordering of $Li^+$ and $Ni^{3+}$. Thus, Ni cations form a triangular lattice sandwiched by Li cation layers. Both cations are coordinated octahedrally by six oxygen anions. The resultant crystal structure is called the delafossite type as shown in Fig. 2(a). Usually, two types of anions with similar ionic sizes order periodically owing to the valence difference. The effective valence of 3+ for the Ni cation suggests a $3d^7$ state as the electron configuration. It can become a Jahn—Teller ion when it is in a low-spin state, as shown in Fig. 2(b), where one electron occupies the $e_g$ state. According to the ligand field theory, five orbital states for the $3d$ orbital split into $t_{2g}$ and $e_g$ states in the octahedron owing to the electron repulsion between the $3d$ orbital electrons and the negatively charged ligand anions [see Fig. 2(b)]. From the quantum chemistry, all the $d$ electrons are in antibonding states, because the bonding states are already occupied by electrons on oxygen anions. This means that $d$ electrons increase the bond lengths between transition metal and oxygen atoms. Two $e_g$ states, which are $x^2-y^2$ and $3z^2-r^2$, have different individual electron distributions. Interestingly, this difference clearly appears as split peaks for the Ni-O bond lengths in the PDF pattern in Fig. 3. Their distances are 1.917(1) and 2.055(1) Å.[24] The peak areas have a ratio of 2:1. This result suggests that the electron occupies the $3z^2-r^2$ state, where only two oxygen ions are elongated in the $z$ direction. On the other hand, four oxygen ions will be elongated in the $x$ and $y$ directions if the $x^2-y^2$ state expands oxygen atoms in the $x$ and $y$ directions. In the case of α-$NaFeO_2$ with $3d^5$ state of $Fe^{3+}$, Five

3$d$-electrons are in the high-spin state, where both $x^2$-$y^2$ and $3z^2$-$r^2$ states are occupied by the electrons. Therefore, all six bond lengths are increased to 2.050(1) Å[26] as shown in Fig. 3. Although there is a small 3$d$ orbital contraction with increasing atomic number, the bond lengths coincide well with the expectation of ligand field theory. The NiO$_6$ octahedra form a triangular lattice in a plane. From the model fittings, the orbital ordering was found to be stabilized as a 120º structure similar to the well-known triangular spin structure, leading to an alternating distorted trimer structure. This distortion induces unidirectional tensile stress in the direction normal to the plane. Because of the stress, the PDF peak width at a long distance becomes broad at low temperatures, suggesting a shorter crystal coherence length with decreasing temperature. This is the opposite tendency to that expected from thermal vibration at higher temperatures. This may be one of the examples of correlated disordered materials.[27]

(a)

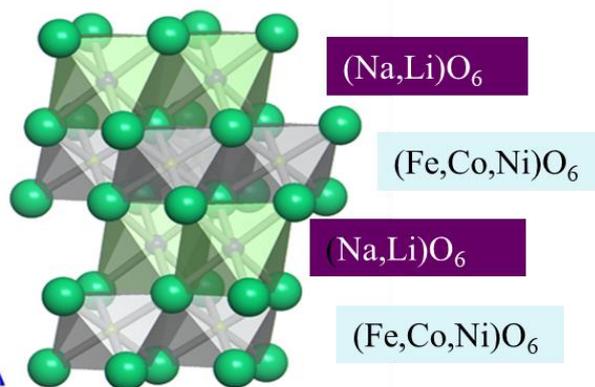

(b)

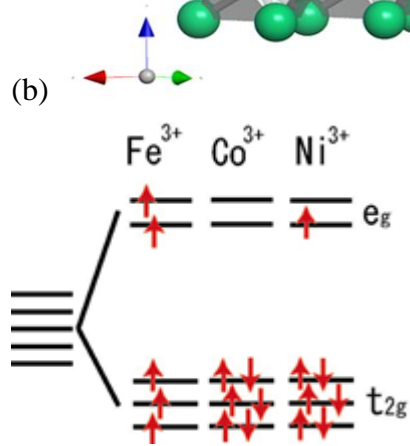

Fig. 2. (Color online) (a) Delafossite crystal structure. (b) $3d$ orbital electron configurations of trivalent Fe, Co, and Ni ions under an octahedral ligand field of six oxygen anions.

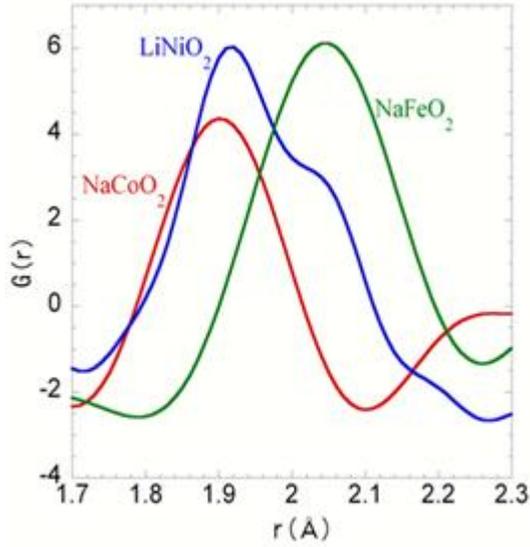

Fig. 3. (Color online) Reduced atomic pair distribution functions (PDFs) in a short range for three compounds, $NaFeO_2$, $Na_{0.7}CoO_2$, and $LiNiO_2$.

One may wonder why there are many insulators that cannot be metallic even after doping with $3d$ transition metal oxides. One of the main reasons is the orbital configuration in the corresponding crystal. For example, in the case of $LiNiO_2$, there is no two-dimensional conducting network of $3d_{3z^2-r^2}$ and $2p$ orbitals because the $2p$ orbitals connected with $3d_{3z^2-r^2}$ are almost orthogonal to each other. In other words, the orbital network is very important in the low-dimensional transition metal oxides for the metal-insulator (MI) transition.

The thermoelectric material γ-$Na_{0.7}CoO_2$ has a similar crystal structure[26] to the delafossite type, where the sodium positions are shifted from the original positions of delafossite. In relation to the Jahn—Teller distortion, γ-$Na_{0.7}CoO_2$ with $d^6$ shows only short bond lengths between cobalt and oxygen atoms, suggesting a low-spin state. The $t_{2g}$ states are fully occupied, whereas the $e_g$ states are empty. The comparison between Co-O and Na-O bonds is intriguing especially in terms of the temperature dependence.

The Co-O bond remains rigid with increasing temperature, whereas the Na-O bond length rapidly increases with increasing temperature. This contrasting behavior can be understood as the characteristic chemical bonds, namely, covalent and ionic bonds. The covalent bond is usually strong. However, the ionic bond easily becomes weak, because the bond strength acceleratedly decreases with increasing the bond length. In this material, the thermal conductivity is closely related to the temperature dependence of the Na-O bond width.[26] Above 600 K, the figure of merit of γ-$Na_{0.7}CoO_2$ is almost entirely determined by the thermal conductivity.[28] This two-dimensional Na diffusion has also been confirmed by $^{23}$Na NMR measurement.[29] This result highlights the importance of thermal vibrational modes in a block layer with little electric conductivity as a means of reducing the total thermal conductivity.

As shown in this section, the short-$r$ PDF analysis gives the local bond lengths of transition metal ions coordinated by oxygen ligands and the chemical bond character in the temperature dependence. The analysis of an ionic conductor shows that the frame structure remains in a solid, whereas the conducting ions become a liquid. Similarly, orbital ordering can be regarded as the solidification of an orbital liquid.[30] Alternatively, the liquid state appears above the ordering temperature. This type of phase transition takes place as a jump in the correlation length, which is very similar to the first-order phase transition of water. $LaMnO_3$-related materials and high-$T_c$ cuprates are also known as Jahn―Teller ion compounds. Because they exhibit MI transitions upon carrier doping, they will be discussed in sect. 3.4 on the Mott transition.

*3.2 Optical recording material*

Optical recording materials are composed of metalloid and basic metal elements in the periodic table. They exhibit both crystal and amorphous phases at room temperature. In general, these elements form an amorphous phase after rapid cooling. This is because short-range covalent bonding becomes dominant in the high-temperature liquid phase due to their characteristic temperature dependences mentioned above. These phases have very different electrical conductivities, resulting in contrasting reflectance. The first phase-change material was $Ge_{10}Si_{12}As_{30}Te_{48}$, discovered by Stanford R. Ovshinsky,[31］ which can be used for electrical switching between high and low electrical conductivities. The crystal and amorphous phases are highly reflective and transparent, respectively. The rapid phase transition between them induced by laser irradiation enables industrial applications as optical recording materials, e.g., DVD and Blu-ray discs. The main element, arsenic, forms a distorted pseudocubic crystal structure.[3］ The electrons are filled in half of the $4p$ orbitals, resulting in Peierls instability. Therefore, the Peierls transition alternately induces short and long bond lengths in the arsenic structure, resulting in a layered structure. GeTe is an isoelectric compound with As. It forms a trigonal structure, which is a distorted NaCl type. At the same time, GeTe can also be amorphous. One of the optical recording materials, $Ge_2Sb_2Te_5$, similarly exhibits crystal and amorphous phase transitions. $Ge_2Sb_2Te_5$ is employed in DVD-RAMs, providing them with a high recording repeatability such as 100,000 times. Depending on the cooling rate, even in the crystal phase, there are two types of crystal structures, namely, cubic and trigonal. The cubic phase has the NaCl-type crystal structure and a silver color. This phase can be prepared only as a thin film. A neutron scattering measurement of these thin films was carried out by collecting powder samples peeled from the films on substrates.

The obtained PDF pattern is shown in Fig. 4. The diffraction pattern was measured at NPDF of LANSCE[16] with high intensity and high $Q$-resolution. The first PDF peak was asymmetric in $r$-space. In the sample, a germanium atom has the largest scattering length for neutrons. For the PDF analysis, all peaks are fitted by the least square fitting method based on the cubic crystal structure with all possible displacements of atoms.[10] Many trials of the initial parameters were required to avoid the local minimum of parameters. As the best result, the asymmetric first peak was successfully reproduced mainly by the germanium displacement[32] with short and long bonds as shown in the bottom panel of Fig. 4. On the other hand, antimony and tellurium atoms are relatively well ordered. By using this local structure model fitting, the $R_{wp}$ factor is reduced from 14.0 to 12.1%. The obtained local structure model is shown in Fig. 5. It is important to note that the cation sites are strongly disordered owing to the large germanium displacement in addition to the random distribution of germanium, antimony, and vacancies, whereas the anion sites are relatively well ordered.[32] Interestingly, the electric conductivity is governed by the hole carrier in the valence band based on the electronic band calculation.[17,32] As shown in Fig. 6, the valence band is mainly composed of the tellurium $5p$ orbital. This suggests that the electric conduction is not disturbed by the disorders at the cation site.[17]

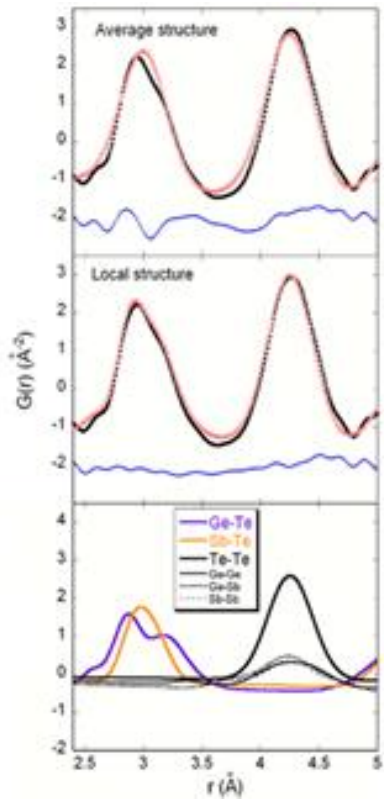

Fig. 4. (Color online) Reduced atomic pair distribution functions (PDFs) of $Ge_2Sb_2Te_5$ cubic phase. The upper panel shows fitting by crystallographic parameters obtained by Rietveld analysis. The middle panel shows a refinement by a local structure model. The difference between the observed PDF and the model PDF is small as shown by the blue line in the panel. The bottom panel shows individual PDFs of the pairs of atoms indicated in the inset. Reprinted with permission from Shamoto et al.[32] (Copyright © *2005 American Institute of Physics*).

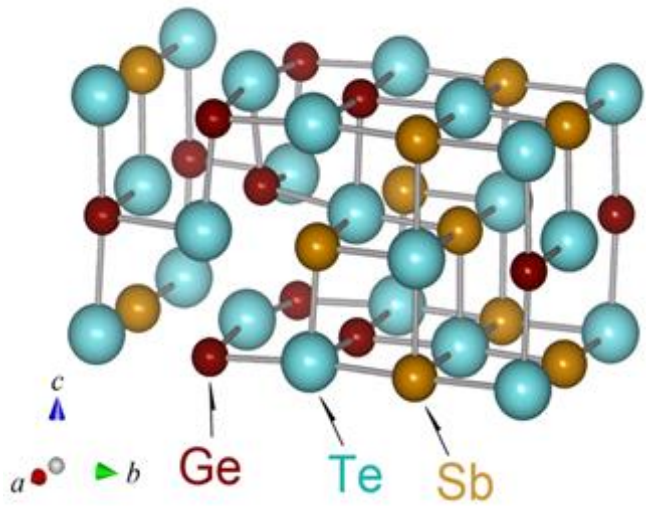

Fig. 5. (Color online) Local structure of $Ge_2Sb_2Te_5$ cubic phase obtained by PDF analysis. Reprinted with permission from Shamoto et al.[17] (Copyright © *2006 The Japan Society of Applied Physics*).

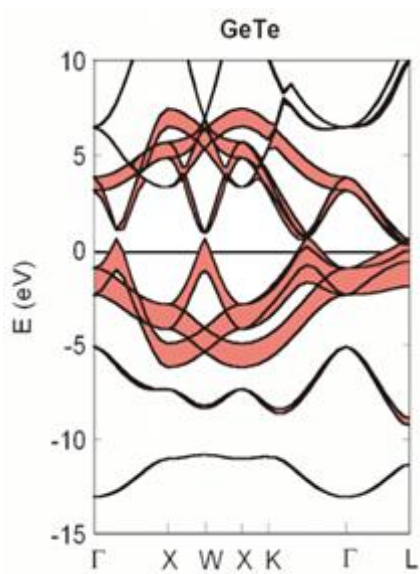

Fig. 6. (Color online) GeTe band structure with isoelectronic structure with $Ge_2Sb_2Te_5$. The width of the bands is proportional to the proportion of the Te $5p$ orbital component. Reprinted with permission from Shamoto et al.[17] (Copyright © *2006 The Japan Society of Applied Physics*).

Regarding the germanium displacement, there can be two sites for a germanium atom in the cubic NaCl phase. One is the Na site, whereas the other is a tetrahedral site coordinated by four Te anions, corresponding to the CuCl-type structure. In the latter case, the germanium atom is tetrahedrally coordinated by tellurium ions. The tetrahedral coordination for elements usually originates from the $sp^3$ hybridization similar to that in a diamond structure. The rapid phase transition is attributed to the site change of germanium atoms.[33] In the ordered NaCl structure, the ionic interaction can have a large influence on the crystal structure because of the long-range interaction. On the other hand, in the liquid (or amorphous) phase, the local covalent bond plays an important role owing to the short-range interaction. There are several merits of this disordered material for optical recording. One of them is the low thermal conductivity,[34] which can reduce the memory area, resulting in high-density memory. On the other hand, it may reduce the entropy change. The latent heat at the first phase transition corresponds to the total heat power for the phase transition. It is proportional to the entropy change. It is 24 J/mol-K for NaCl at the melting point. A similar compound, $GeSb_2Te_4$, shows an entropy change of 13.7 J/mol-K from amorphous to cubic phases at $T=130\ °C$.[35] This is very close to half of the entropy change of NaCl. It is also interesting to note that the ionic conductor α-AgI shows an entropy change of 11.3 J/mol-K upon melting. The entropy change from β-AgI (poor ionic conductor) to α-AgI (excellent ionic conductor) is 14.5 J/mol-K at $T = 146$ ºC. Because of these entropy changes, the ionic conductor α-AgI is regarded as a half-melted material. A similar discussion can be applied to this crystal phase of the optical recording material. Therefore, these crystal phases of optical recording materials can be called half amorphous phases.

Among the optical recording processes, the writing of a cubic crystalline phase is the most time-consuming part because it includes a crystallization time. It is important to reduce the crystallization time for rapid recording. The crystallization process has been studied for various cases.[36-38] The crystal growth rate is proportional to the fraction of sites $f$ at the interface, the diffusion constant $D$", and the free energy change $\Delta G$ associated with crystallization in the high-temperature limit. In addition, it is inversely proportional to the distance $\lambda$ between crystalline and amorphous atomic sites.[38] The local atomic structure of the amorphous phase is found to have a rocksalt crystal-like topology similar to that of the crystalline phase,[39] suggesting a short distance between the crystalline and amorphous atomic sites. This is considered to be why the rapid phase change is achieved in this phase. The reduced $\Delta G$ for the first-order phase transition caused by the randomness in a crystalline material shortens the laser irradiation time owing to the small latent heat, although the nucleation speed is decreased.

*3.3 Negative thermal expansion*

One of the practical applications of itinerant magnets is negative thermal expansion, as discovered in the Invar alloy.[40] Controlling the thermal expansion is increasingly important for the precise control of small devices. The traditional mechanism used is the magnetoelastic effect. Upon magnetic ordering, the volumes of intermetallic alloys increase with decreasing temperature. In the case of transition metals, the electronic $3d$ bands are composed of a bonding-orbital band with a large band width and an antibonding band with a small band width. If the narrow antibonding band at a high energy is half-filled, the system will have spin splitting under $UN(E_F)>1$ as a Stoner instability[41] with on-site Coulomb interaction $U$ and electronic density of states at the Fermi energy $N(E_F)$. By the spin splitting, electrons are transferred from the down-spin band to the up-spin band, where the total energy is reduced, leading to magnetic ordering. The transfer integral between neighboring orbitals is obtained on the basis of the second perturbation, resulting in the electronic band structure. The paramagnetic metallic state exhibits Curie—Weiss behavior in the temperature dependence of magnetic susceptibility, suggesting a localized spin nature.[42] Once the spin correlation between the two neighboring sites develops, the transfer integral becomes strongly spin-dependent. The local-band theory for a ferromagnetic state has been discussed in some studies.[43,44] The magnetic moment size of 3$d$ electrons can be very sensitive to the coordinated anion position near a critical point, leading to a strong spin-lattice coupling.[45-47]

Regarding negative thermal expansion, the first type of material with this property is intermetallic alloys such as iron-nickel alloys and $Fe_3Pt$.[48] The second type of material is cluster network materials. For example, a material may have a cluster network of tetrahedra and/or octahedra connected by apical oxygens. If the structure network has sufficient free space for the rotational vibration mode, a large vibration amplitude may reduce the bond length between transition metal atoms. This type of material usually shows a wide temperature range of negative thermal expansion, such as $ZrW_2O_8$[49] and $LiAlSiO_4$.[50] The third type of material with negative thermal expansion is the charge ordering material $(Bi_{1-x}La_x)NiO_3$.[51] The nominal charge at Bi and Ni cations is 3+. After a charge is transferred from Bi to Ni sites, $Bi^{4+}$ becomes the valence skip cation of $Bi^{3+}$ and $Bi^{5+}$. This charge redistribution changes the total volume dramatically, leading to a colossal negative thermal expansion near room temperature. Because every negative thermal expansion is related to the local structure, PDF analysis can play an important role in revealing the mechanism.

Antiperovskite materials such as $Mn_3GaN$[52] exhibit a large negative thermal expansion as the first phase transition. In some applications, a wide temperature range of negative thermal expansion is required. By Ge doping at the Cu site in $Mn_3CuN$, the phase transition was transformed from the first to the second phase transition, resulting in a wide temperature range of negative thermal expansion.[53] On the other hand, in the case of Ga doping at the Cu site, the first-order phase transition remained up to the end member $Mn_3GaN$, although the transition temperature increases to near room temperature.[54] The transitions clearly appear in the temperature dependences of lattice parameters and magnetic Bragg peak intensities[55] as shown in Figs. 7 and 8. The results are consistent with the magnetovolume mechanism. The difference between Ge and Ga was a puzzle, because their crystal structures are very similar to each other including their lattice parameters. Although the transition change to the second order was achieved only by Ge doping, it was difficult to understand the mechanism from Rietveld analysis. The magnetic structure is an antiferromagnetically ordered structure in $\Gamma^{5g}$ with a $120^o$ structure in the cubic (111) plane (Fig. 9),[55] although the magnetic moments are parallel for the pairs of Mn-N-Mn linear bonds.[56] The magnetic structure in a Mn triangle of the (111) plane reminds us of a large volume change of about 5% at the Neel temperature $T_N$ in the Laves phase of $YMn_2$,[57] where the Mn site forms a well-known frustrated pyrochlore spin lattice. This suggests that the magnetic ordering from the frustrated spin paramagnet to AF has a large magnetovolume effect. To study the thermal expansion mechanism, the local structure was studied as shown in Fig. 9.[58] In this material, the local static rotation of the Mn$_6$N octahedron was discovered, which could not be observed in the standard crystallographic Rietveld analysis because of the random rotation. The rotational angle increased with increasing Ge content. Although the discovered angle was about 4$^o$, this small rotation is expected to reduce the electric band width, because of a slight change in the lattice parameters. The critical angle between the phase transitions is estimated to be about 3$^o$ as shown in Fig. 11. At the end member of Mn$_3$GeN, the octahedral rotation was frozen below $T$= 540 K, resulting in the lowering of the crystal symmetry of $I$4/$mcm$. From our PDF analysis, a large octahedral rotation of above 8$^o$ was discovered in Mn$_3$GeN even above the phase transition. The rotation angle increases slightly with increasing temperature as shown in Fig. 11. The small rotational angle of Mn$_3$Cu$_{1-x}$Ge$_x$N can be regarded as a freezing of the phase transition to the tetragonal phase. This freezing phenomenon is exactly the same as that observed in a ferroelectric relaxor. The specific local lattice distortion around the Ge atom was also discovered by EXAFS measurement.[59] The Ge atom seems to preferentially have a lower symmetry in general, as also shown in the previous section for optical recording material.

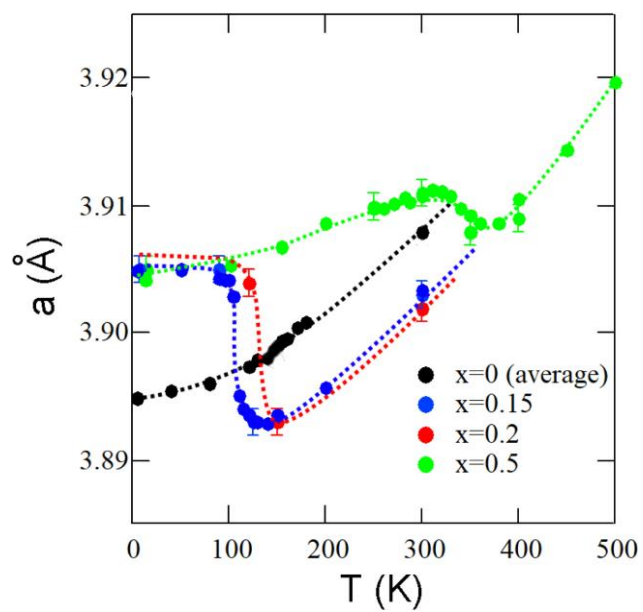

Fig. 7. (Color online) Lattice parameters of $Mn_3(Cu_{1-x}Ge_x)N$ as a function of temperature.[56]

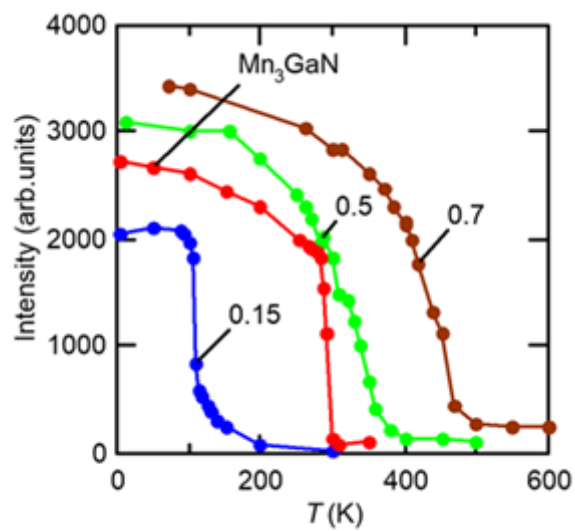

Fig. 8. (Color online) Magnetic Bragg peak intensities of $Mn_3(Cu_{1-x}Ga_x)N$ as a function of temperature.[56]

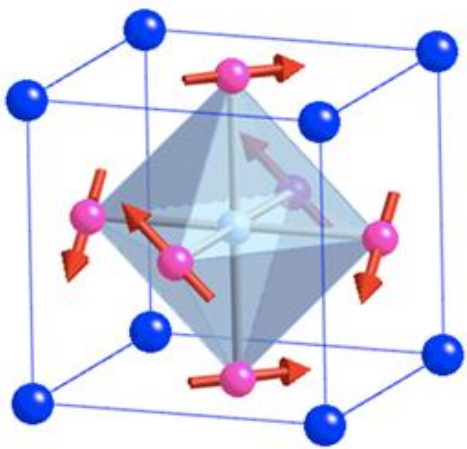

Fig. 9. (Color online) Crystal structure of antiperovskite with $\Gamma^{5g}$ magnetic structure.

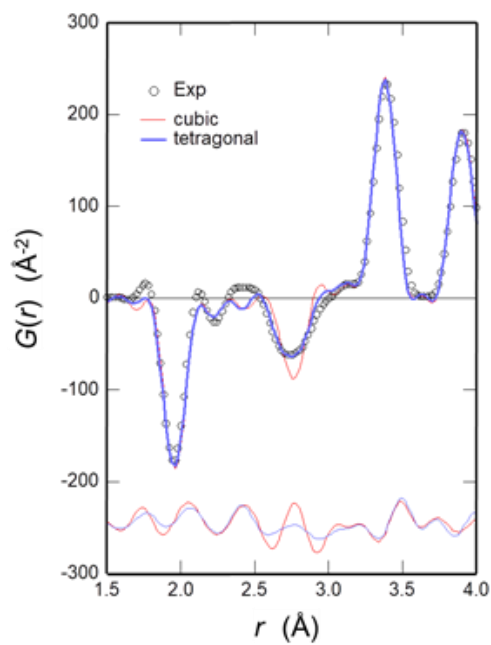

Fig. 10. (Color online) Reduced atomic pair distribution function of Mn$_3$(Cu$_{0.5}$Ge$_{0.5}$)N in the short $r$-range below 4 Å.[59)] The nonrotational distorted model PDF is also shown for comparison. The difference shown at the bottom supports the validity of the rotational mode at around 2.8 Å.

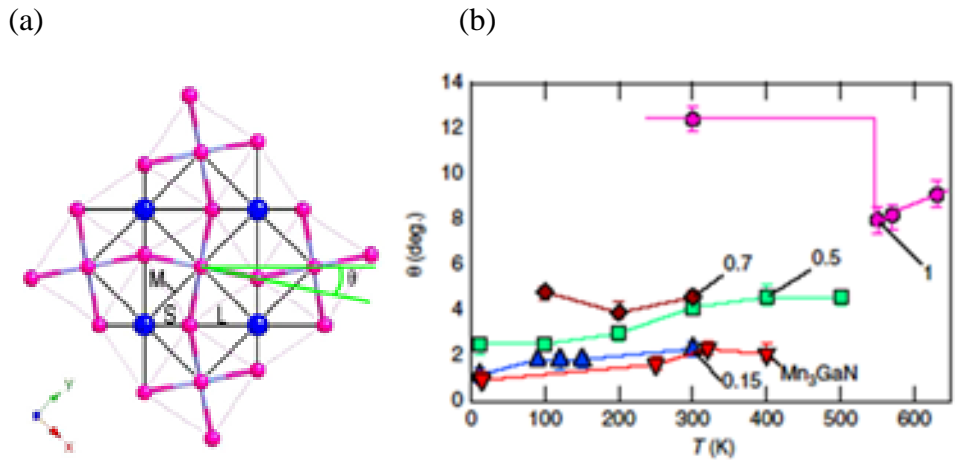

Fig. 11. (Color online) (a) $Mn_3N$ octahedral rotation model with an angle of $\theta$. (b) Temperature dependence of the rotational angle estimated by PDF fittings for various materials. Reprinted with permission from Iikubo et al.[59] (*Copyright © 2008 The American Physical Society*).

Here, there are two important aspects of the negative thermal expansion. One is the size of the volume change. The other is the temperature width for the negative thermal expansion. A large volume jump can be expected at a first-order phase transition. The wide temperature range requires the second-order phase transition if it is related to the magnetovolume effect. According to a local structure study on $Mn_3(Cu_{1-x}Ge_x)N$, the latter issue was important to change the phase transition. It can be controlled by changing the band width. At the same time, the local octahedral rotation is frozen similarly to a ferroelectric relaxor, which is hindered in a standard crystallographic Rietveld analysis.

A similar change from the first- to second-order phase transition has been observed in the $RCo_2$ system (R: rare-earth elements) in the Laves phase, where the band width decreases with increasing ionic radius of R.[60] It is intriguing to note that the first-order phase transition takes place near a critical point of the magnetic ordering, similarly to the Mott transition.[61] By increasing the band width, the second-order phase transition changes to the first-order phase transition, where the energy gain by magnetic ordering must be reduced owing to the band width broadening. The band width also decreases with increasing rotational angle in the case of $RNiO_3$.[61,62] According to a systematic study, with decreasing rare-earth ionic radius, the tolerance factor of the perovskite structure decreases, which corresponds to the increase in the rotational angle. The MI transition temperature increases with increasing rotational angle. This can be regarded as an increase in $U/W$, where $U$ is the on-site Coulomb interaction and $W$ is the band width. This is because the band width $W$ decreases with increasing rotational angle. As a result, a local octahedral rotation plays an important role in changing the phase transition type. The superconducting anomaly of $La_{2-x-y}Nd_ySr_xCuO_4$ at $x=1/8$ also shows a critical tilt angle for the static stripe phase of $3.6°$.[63] The present critical angle of $3°$ is similar to this value. Because the octahedral rotation can change the band width, as shown here, it should be possible to couple with the Fermi surface instability as Fermi surface nesting. For example, the mixing of two phases with different octahedral rotation angles or bond lengths may result in a gap at the Fermi surface similar to a charge density wave. It may be called a transfer integral wave due to the modulation of transfers between neighboring atoms.[64] This discussion may be extended to include the local spin correlation dependence in the hopping term.

According to the mechanism of negative thermal expansion, the volume change is related to the entropy change. From the Clausius—Clapeyron equation,[65] a small-volume phase with a high entropy becomes more stable at high temperatures when the entropy of a high-pressure phase with a small molar volume is larger than the entropy of a low-pressure phase with a large molar volume.[66] According to this scenario, the entropy difference is proportional to the volume change. The large entropy change together with the volume change is applied to the barocaloric effect.[67] The temperature is lowered with a rapid release of hydrostatic pressure on $Mn_3GaN$. It reaches 5 K for a depressurization of 93 MPa. The magnetocaloric effect is also similarly related to the entropy change. It can be enhanced by tuning the structural instability at the magnetic phase transition.[68] Therefore, any transition such as MI transition[69] may contribute to the volume change accompanied by the entropy difference.

*3.4 Mott transition and electron-lattice coupling*

The Mott transition is accompanied by the MI transition, which can be induced by carrier doping or pressure. The change can be dramatic because of the full localization of electrons. At the same time, the bond lengths in the conducting network should change accordingly. In the case of a small amount of carrier doping, the change is not so large but has been detected using existing neutron spectrometers.

In the case of $RNiO_3$, the magnetovolume effect of negative thermal expansion was observed as a first-order phase transition.[61,62] At the same time, the resistivity also shows a sudden jump from metal to insulator, with a sudden decrease in the paramagnetic fraction. This result is consistent with the μSR result,[61] where a phase separation is clearly observed as a magnetic ordered volume fraction near the Mott quantum critical point. The Ni ion is expected to have a valence of +3 in the octahedron of oxygen anions, resulting in one electron in the $e_g$ state as a low-spin state as observed in $LiNiO_2$. The high-$T_c$ cuprate $La_{2-x}Sr_xCuO_4$ also exhibits a similar phenomenon in the metallic region. The first-nearest-neighbor peak between Cu and O in the PDF has intrinsic broadening in the superconducting dome, suggesting charge inhomogeneity or phase separation.[70)] The PDF peak width becomes the largest at the highest-$T_c$ hole concentration near the quantum critical point. In addition, the PDF of the $x=0.1$ sample is well described by an equal mixture of the PDFs of non-superconducting $x=0$ and $x=0.25$ samples.[71)] The Cu-O bond length change from $x=0$ to $x=0.25$ samples is about 0.023 Å. This leads to a large bond length change of about 0.09 Å per hole carrier, which is roughly the same as the C-C bond length change mentioned above. According to an EXAFS measurement,[72)] phase separation is also discovered in the $CuO_2$ plane of optimally doped $La_{1.85}Sr_{0.15}CuO_4$, where the largest bond variation reaches 0.08 Å. The in-plane Cu-O bond length decreases owing to the hole doping because the Cu-O antibonding electrons are removed. This bond length change can also be explained by the cluster model of $CuO_6$.[5)] The electronic state is determined in a mixed state of $d^9$ and $d^{10}L$ ($L$: ligand hole), where the energy difference between $d^9$ and $d^{10}L$ is called a charge transfer gap, $\Delta$. Upon hole doping, the Cu-O distance decreases due to the hybridization among $d^8$, $d^9L$, and $d^{10}L^2$, where the energy differences are $U-\Delta$ ($d^8$ and $d^9L$) and $\Delta$ ($d^9L$ and $d^{10}L^2$). In the case of the three-dimensional Mott system of $RNiO_3$, the quantum critical point is defined at the transition point from AF ordering to the paramagnetic state at $T=0$ K.[61)] Phase separation is observed in the AF phase.[61)] In the high-$T_c$ cuprate $La_{2-x}Sr_xCuO_4$, peak broadening due to phase separation is observed in the metallic state under a pseudo-gap transition.[70)] The analogy of the phase diagram of the $La_{2-x}Sr_xCuO_4$ system with the $RNiO_3$ system suggests that the magnetic ordering of the high-$T_c$ cuprate $La_{2-x}Sr_xCuO_4$ may be suppressed owing to the low dimensionality in addition to the mobile hole carriers. Note that the $(\pi, \pi)$ AF ordering is consistent with the Fermi surface wave vector for the Fermi surface nesting. In addition, the effective carrier numbers are largely reduced below the pseudo-gap temperature,[73] which can also be attributed to the phase separation in the region. The two phases (AF insulating phase and metallic phase) may be fluctuating in time and space. The average local structure with a broad peak width is obtained at an instantaneous atomic correlation because of the wide energy integration. To estimate the time scale, it is necessary to measure the energy or time dependence, for example, through the dynamical PDF method.[74]

Colossal magnetoresistance manganite[75] is one of the good examples for learning the relationship between local structure and physical properties.[76,77] $Mn^{3+}$ ($3d^4$) is a Jahn—Teller ion in the high-spin state. When a hole is doped in $LaMnO_3$, the lattice distortion due to the Jahn—Teller effect may disappear. At 300 K, however, the largest bond length is found to be kept in a metallic sample with up to 20% Ca doping. For the parent compound $LaMnO_3$, the distortion appears below the phase transition temperature of 750 K from pseudocubic to rhombohedral phases due to orbital ordering.[30] Above the phase transition temperature, the local structure can be ascribed to an orbital liquid, where only a short correlation length of the lattice distortion survives even at 1100 K.[30] The local lattice distortion in a short range, which may be called a polaron,[76] remains with the same amplitude. The temperature dependence of the correlation length exhibits a jump at the structural phase transition temperature.[30] The first-order phase transition can be described by the jump of the correlation length, which corresponds to the entropy jump. This is exactly the same as the first-order solid-liquid phase transition, where only the jump of the atomic pair correlation length is observed structurally.

According to a systematic local structure study on the $La_{1-x}Ca_xMnO_3$ phase diagram, a metallic state at low temperatures is realized by decreasing the Jahn—Teller distorted long-bond-length in the PDF.[78] The electronic conduction may be described as a conductive network among $MnO_6$ clusters, where the nucleation of the distorted $MnO_6$ cluster even in a short range degrades the electric conductivity. The existence of the distorted cluster results in the insulating phase, where the electrons (polarons) cannot hop between the distorted clusters. The hopping probability may be expressed by the activation type with a gap energy because the orbital ordering prevents the electrons from hopping between these clusters. The largest bond length change reaches about 0.3 Å/e. A large jump of about 0.1 Å (from 2.06 to 1.97 Å) is observed at the MI transition at $x$~0.22, accompanied by a magnetic transition from AF to ferromagnetic phases. The 3$d$ band of the Mott insulator is fully polarized. Note that the phase separation in the doped $LaMnO_3$ system is also observed near the tricritical point in the Ca- or Sr-doped $La_{1-y}Pr_yMnO_3$.[79,80]

Because of these large bond length changes, a large electron-lattice coupling is expected. The electron—phonon interaction of a high-$T_c$ copper oxide is estimated on the basis of the tight-binding electronic bands without the Coulomb interaction.[81] It includes the bonding and antibonding orbital effects. However, no on-site Coulomb repulsion effect or spin correlation effect on the bond lengths is included. The electron correlation effect on the electron—phonon interaction has been discussed.[82-84] The important point is that the electron—electron interaction is repulsive for electrons, whereas the electron—phonon interaction is attractive for electrons. Because these interactions are mutually exclusive, the phase boundary becomes a first-order transition.[82] In the phase diagram, separation of the charge density wave phase (with strong electron—phonon interaction) and AF phase (with strong electron—electron interaction) appears.[81] Although the calculation has been carried out only with a small four-site-four-electron model, similar phenomena have been observed in a high-$T_c$ cuprate phase diagram.[85] The observed bond lengths vary largely and inhomogeneously upon carrier doping, suggesting large electron—lattice coupling, especially near the critical point. The complementary use of real-space and reciprocal-space analyses may play an important role in exploring the Mott effect on the bond lengths, similarly to other methods.[86] Further study including the energy dependence and time evolution is expected to reveal the intrinsic phenomena.

## 4. Conclusions

The local structural point of view gives us a deeper understanding of materials, as shown here. The local structure represents the electronic state as a chemical bond. For example, the chemical bond has a characteristic temperature dependence. The covalent bond has a short-range correlation whereas the Madelung potential can be regarded as a relatively long range correlation because of the $r$-dependence of potentials. In amorphous and liquid phases the bond correlation decreases quickly with increasing distance owing to the large randomness, where covalent bonding plays an important role. A similar transition has been observed locally in a crystalline solid. The PDF obtained from a powder sample shows the average summation of instantaneous atomic pair correlations in time and space. Although it is still a powerful tool for observing the complex atomic correlations, powder averaging may lose the individual long-range PDF information even in a high-$Q$ and high-$r$ resolution PDF. In particular, in multicomponent materials, it becomes difficult to analyze the PDF peaks with the specific atomic correlations because of the broadening of the PDF peaks. In such a case, it requires the PDFs of specific elements, which can be obtained by using isotopes in

neutron scattering and the resonant energy in X-ray scattering.[87]

In order to detect specific small structural changes on the surface or by doping, differential PDF[20] analysis can be applied. Depending on the ratio of the amorphous part in your material, the analysis method may vary. The present PDF patterns are analyzed by PDFFIT,[88] which is easy to use for crystalline materials with small disorder from crystallographic sites. On the other hand, largely disordered materials require reverse Monte Carlo modeling[89]. This preference is based on the fitting procedures. The former program is based on a real-space nonlinear least square fitting method for atomic site refinement. The latter program is based on a reverse Monte Carlo method with the random trial movement of atoms in a relatively large model. The larger disorder may include a larger ambiguity for the model structure owing to the intrinsic ambiguity, which usually requires multiple measurements. Here, only the PDF of one-dimensional data as a function of $Q$ was presented. Diffuse scattering is often observed to have some structure such as streaks, owing to anisotropic local atomic pair correlation. This type of local atomic pair correlation is discussed as a correlated disorder.[27] For example, water molecules have intrinsic disorder known as an ice rule. Similarly, a frustrated spin system could also have the same type of disorder as a result of a spin ice rule. 3D PDF analysis may play an important role in correlated disorder materials. In addition, long-range correlated disorder could be analyzed by the 3D PDF method with much information. Some local magnetic structures are solved by magnetic PDF[90,91] and reverse Monte Carlo methods.[92]

These newly developed techniques including dynamical PDF will help us explore the frontier of disordered functional crystalline solids by using high intensity and high $Q$-resolution total elastic/inelastic scattering spectrometers at pulsed neutron facilities.

**Acknowledgments**

I would like to thank all my collaborators, K. Kodama, S. Iikubo, T. Taguchi, K. Takenaka, H. Takagi, M. Takigawa, N. Yamada, T. Matsunaga, Th. Proffen, J. W. Richardson, H. Koyanaka, K. Takeuchi, S. Kohara, C.-K. Loong, K. Ikeda, T. Otomo, Y. Hasegawa, T. Kajitani, M. Sato, J.-H. Chung, and T. Egami. The Mott transition effect and spin-lattice coupling in the present discussion were inspired by the discussions with Y. Uemura, A. Fujimori, and T. Egami.